\documentclass[prl,twocolumn,showpacs,preprintnumbers,amsmath,amssymb]{revtex4}

\usepackage{graphicx}
\usepackage{amssymb,amsmath}
\usepackage{dcolumn}
\usepackage{bm}
\usepackage{color}

\begin{document}

\title{Phonon-mediated Josephson oscillations in excitonic and polaritonic condensates}

\author{E. B. Magnusson}
\affiliation{Science Institute, University of Iceland, Dunhagi-3,
IS-107, Reykjavik, Iceland}

\author{H. Flayac}
\affiliation{Clermont Universit\'{e}, Universit\'{e} Blaise Pascal, LASMEA, BP10448, 63000 Clermont-Ferrand, France}
\affiliation{CNRS, UMR6602, LASMEA, 63177 Aubi\`{e}re, France}

\author{G. Malpuech}
\affiliation{Clermont Universit\'{e}, Universit\'{e} Blaise Pascal, LASMEA, BP10448, 63000 Clermont-Ferrand, France}
\affiliation{CNRS, UMR6602, LASMEA, 63177 Aubi\`{e}re, France}

\author{I.A. Shelykh}
\affiliation{Science Institute, University of Iceland, Dunhagi-3,
IS-107, Reykjavik, Iceland}
\affiliation{International Institute for Physics, Av. Odilon Gomes de Lima, 1722, CEP: 59078-400, Capim Macio, Natal- RN,
Brazil}

\date{\today}

\begin{abstract}
We analyze theoretically the role of the exciton-phonon interactions in phenomena related to the Josephson
effect between two spatially separated exciton and exciton-polariton condensates.  We consider the role of the dephasing introduced by phonons in such phenomena as Josephson tunneling, self-trapping and spontaneous polarization separation. In the regime of cw pumping we find a remarkable bistability effect arising from exciton- exciton interactions as well as regimes of self- sustained regular and chaotic oscillations.
\end{abstract}

\maketitle

\section{Introduction}

Collective phenomena lie beyond many remarkable effects in condensed matter physics. One of their famous manifestations is Josephson effect \cite{Josephson}, which was first predicted to occur between superconductors separated by a thin dielectric layer. Due to the build-up of the macroscopic phase coherence resulting in the appearance of an order parameter (playing a role of macroscopic wavefunction of the Cooper pairs) $\Psi(\textbf{r},t)=\sqrt{\rho}e^{-i\phi}$, a tunnel current appears between superconduction regions proportional to the sine of the phase difference between them \cite{Legett}:
\begin{equation}
I=I_0\sin\Delta\phi
\end{equation}
where $I_0$ is a constant depending on the properties of the junction.

Later on, it was proposed that similar phenomena can be observed using liquid Helium \cite{Pereverzev} and cold atoms \cite{Albiez}, where the appearance of a macroscopic wavefunction accompanies the transition towards superfluid and BEC states \cite{Anderson} respectively. In this last case, Josephson effect can take place between two spatially separated BEC of atoms, weakly coupled trough a barrier. The situation there can demonstrate new physical phenomena with respect to the original junctions between superconductors, as interactions between the tunneling particles play a major role and can lead to remarkable nonlinear effects in the Josephson dynamics. These effects are the anharmonicity of the Josephson oscillations \cite{Jack} and macroscopic self-trapping in the case, when the initial imbalance between the two condensates exceeds some critical value \cite{Smerzi}. The serious disadvantage of the cold atom systems is that corresponding critical temperatures are extremely small (usually in the nano-Kelvin range) and thus any experimental investigations in the field become difficult. Besides, low critical temperatures rule out any possibility of using the system for practical applications.

On the other hand, in the field of condensed matter physics, various candidates were proposed for the realization of BEC with critical temperatures orders of magnitudes higher then those of cold atoms. The formation of  exciton condensates in bulk semiconductors was theoretically predicted more than 40 years ago \cite{Keldysh}, but appeared to be difficult to realize experimentally. Since then, other solid- state systems were proposed for the achievement of high-temperature BEC, including Quantum Hall bilayers
\cite{Eisenstein}, magnons \cite{Democritov}, undirect excitons \cite{Timofeev,Butov} and
cavity exciton-polaritons \cite{KasprzakNature,Balili,BaumbergPRL2008}. The latter two systems will be
in the focus of the present paper.

Spatially indirect excitons have been widely studied both experimentally and
theoretically in recent years (see Ref.\onlinecite{Butov} for a review). For such particles, electrons and holes are localized in parallel coupled 2D layers. Their wave functions show a very little overlap and consequently, indirect excitons have a very long lifetime (tens of milliseconds), and can be treated as metastable particles. Superfluid behavior of a system of indirect excitons has been predicted by Lozovik and Yudson more then 30 years ago \cite{Lozovik} and subsequent theoretical \cite{Shevchenko,Zhu,Berman} and experimental \cite{Larionov,Snoke2002,Butov2002} studies have
suggested that this should be manifested in a series of remarkable effects, including persistent currents and Josephson- related phenomena.

Exciton- polaritons are elementary excitations of quantum
microcavities in the strong coupling regime. They have a hybrid nature and represent a combination of
quantum well excitons and cavity photons, which gives to them a number of peculiar properties distinguishing them from other quasiparticles.  Due to the presence of the photonic
component, the effective mass of exciton-
polaritons is 4-5 orders of magnitude smaller than the one of free electron, which makes them the lightest quasiparticles in condensed matter systems \cite{KavokinBook}.  The possibility for the photons to leave the system through dielectric Bragg mirrors forming the cavity together with the processes of non-radiative recombination of the excitons makes lifetime $\tau$ of cavity polaritons finite (longest lifetimes reported up to now were in a range of tens of picoseconds \cite{Wertz}). The presence of the excitonic component makes
possible effective polariton-polariton interactions. These factors
are crucial for polariton BEC, whose critical temperature was shown to be surprisingly high, 20K in CdTe cavities \cite{KasprzakNature} and up to room temperature in GaN cavities \cite{Christopoulos, BaumbergPRL2008}. One should note, however, that differently from the cases of cold atoms and indirect excitons, and depending on the experimental configuration, polariton condensation can be a strongly out- of equilibrium process governed by relaxation kinetics and which cannot be described in this regime by conventional thermodynamic BEC \cite{Kasprzak2008,Krizhanovskii2009,Maragkou, Levrat2010}.

An important peculiarity of the polariton system lies in its spin
structure \cite{DarkSpinRemark}: being formed by bright heavy-hole excitons, the lowest
energy polariton state has two allowed spin projections on the
structure growth axis ($\pm1$), corresponding to the right and left
circular polarizations of the counterpart photons. The states having
other spin projections are split-off in energy and normally can be
neglected while considering polariton dynamics.
Moreover, due to the effects of exchange, the inter-particle
interactions are strongly spin-anisotropic \cite{Ciuti,Combescot,Ostatnicky}, the strength of the interaction between polaritons of same circular polarization being order of magnitude stronger that for polaritons with opposite spin polarizations \cite{Kirill}. The combination of their finite lifetime and of spin- related phenomena makes that exciton and polariton condensates behave differently from atomic condensates or superfluids even in the thermodynamic limit \cite{ShelykhPRL,ShelykhSST}.

The possibilities, recently demonstrated of engineering spatial confinement for excitons and exciton-polaritons \cite{Hammack,Balili,BaliliAPL,Daif,Wertzarxiv} opens a way to the investigation of the Josephson effect based on the tunneling between two spatially separated condensates of these particles \cite{Wouters,Savona,Shelykh2008,Solnyshkov2009,Read2010}. As mentioned previously, Josephson effect for excitons and cavity polaritons have several important differences from those for superconductors and cold atomic BECs.

First, the inter-particle interactions play by far a more important role here, leading to the anharmonicity of the Josephson oscillations and to the self-trapping effect \cite{Jack,Smerzi}.

Second, the presence of the polarization (spin) degree of freedom, combined with spin-anisotropy of polariton-polariton interactions gives rise to a much richer and original phenomenology, including spontaneous polarization separation in the real space\cite{Shelykh2008,Read2010}.

Third, due to their short lifetime cavity polaritons (not indirect excitons) cannot be considered as metastable particles, and effects of pump and decay should be accounted for while considering Josephson- related phenomena. Due to the strong polariton- polariton interactions one can expect that effects of bistability \cite{Tredicucci,Baas,Gippius2004} and multistability \cite{Gippius2007,Liew2008,Naturematerials2010} can play an important role.

Fourth, indirect excitons and cavity polaritons efficiently interact with phonons, which play the role of a source of decoherence and can also affect the tunneling rates \cite{Caldeira}

In the present work we focus on those last two effects. The paper is organized as follows. In the Section II we give a description of the model of Josephson tunneling in a system of coupled excitonic or polaritonic condensates. This model accounts for 1. The exciont/polariton spin structure 2. Spin- anisotropic particle- particle interactions 3. Interactions with acoustic phonons 4. Pump and decay of the particles. In the Section III we present the results for the cases of both indirect excitons (infinite lifetime) and cavity polaritons (finite lifetime). We discuss in detail the effect of the interaction with phonons on the Josephson oscillations and on the self trapping for the case of indirect excitons  and consider the role of multistability in polaritonic system with pump and decay. Conclusions summarize the results of the work.

\section{The Model}

\begin{figure}
\includegraphics[width=1.0\linewidth]{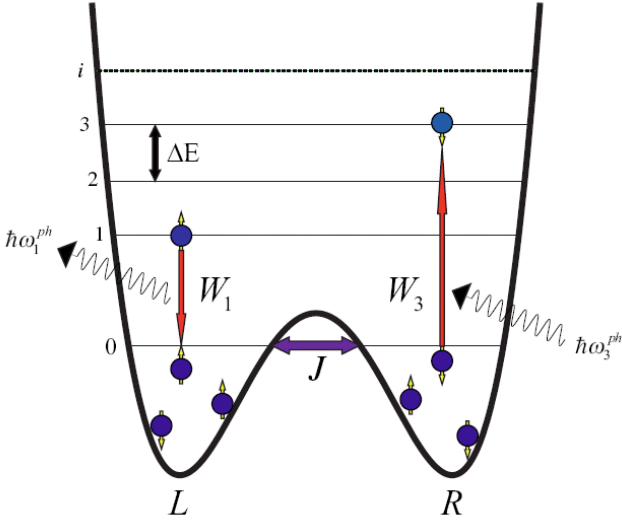}
\caption{Geometry of the system. Spinor polaritons are confined in two traps coupled with each other by coherent tunneling and with delocalized excited states by energy conserving phonon- assisted processes.}
\label{scheme}
\end{figure}

The system we analyze is schematically represented on the Fig.\ref{scheme}. We consider two spatially separated excitonic or polaritonic condensates located in a pair of coupled traps. Each trap contains a single confined discrete level. These two localized states are weakly coupled the one to another by a tunnel constant $J$. This tunneling process can give rise to a coherent oscillation between the right and the left well, which can be described as a bosonic Josephson effect (extrinsic Josephson effect in terms of Ref.\onlinecite{Shelykh2008}). The particles have two spin projections, corresponding to the right ($\sigma_+$) and left-circular ($\sigma_-$) polarizations of the counterpart photons. Due to the structural asymmetry, one can have a  coherent exchange of particles between the condensates with opposite polarizations \cite{Aleiner,Klopotowsky}, referred to as intrinsic Josephson effect \cite{Shelykh2008}. Besides, the system contains excited delocalized levels, which are coupled with localized states in the traps via processes involving acoustic phonons which contribute to the exchange of the particles between the traps and introduce a source of decoherence to the system. An alternative process of transfer of particles from one well to another and based on polariton-polariton scattering could be envisaged. This will involve the simultaneous scattering of one condensed particle to a reservoir state and the scattering of a reservoir particle to the other condensed state. This type of mechanism can lead either to decoherence or to the enhancement of the Josephson coupling constant depending on the coherence degree of the reservoir state \cite{Trujillo}. The description of this mechanism is beyond the scope of the present paper and will be the topic of a separate publication.

The model Hamiltonian is thus separated into two parts:
\begin{equation}\label{H}
    H = {H_{co}} + {H_{deco}}
\end{equation}

The coherent part:
\begin{equation}\label{Hco}
{H_{co}} = {H_0} + {H_J} + {H_\Omega } + {H_{pol - pol}}
\end{equation}
with:
\begin{eqnarray}
{H_0} &=& {\varepsilon _0}\left( {a_{L \uparrow }^\dag {a_{L \uparrow }} + a_{L \downarrow }^\dag {a_{L \downarrow }} + a_{R \uparrow }^\dag {a_{R \uparrow }} + a_{R \downarrow }^\dag {a_{R \downarrow }}} \right)\\
{H_J} &=& J\left( {a_{L \uparrow }^\dag {a_{R \uparrow }} + a_{R \uparrow }^\dag {a_{L \uparrow }} + a_{L \downarrow }^\dag {a_{R \downarrow }} + a_{R \downarrow }^\dag {a_{L \downarrow }}} \right)\\
{H_\Omega } &=& \Omega \left( {a_{L \uparrow }^\dag {a_{L \downarrow }} + a_{L \downarrow }^\dag {a_{L \uparrow }} + a_{R \uparrow }^\dag {a_{R \downarrow }} + a_{R \downarrow }^\dag {a_{R \uparrow }}} \right)
\end{eqnarray}
\begin{equation}\label{Hpolpol}
{H_{pol - pol}} = {H_{ \uparrow  \uparrow }} + {H_{ \uparrow  \downarrow }}
\end{equation}
\begin{eqnarray}
{H_{ \uparrow  \uparrow }} &=& \frac{{{U_1}}}{2}\left( \begin{array}{l}
 a_{L \uparrow }^\dag a_{L \uparrow }^\dag {a_{L \uparrow }}{a_{L \uparrow }} + a_{R \uparrow }^\dag a_{R \uparrow }^\dag {a_{R \uparrow }}{a_{R \uparrow }} \\
  + a_{L \downarrow }^\dag a_{L \downarrow }^\dag {a_{L \downarrow }}{a_{L \downarrow }} + a_{R \downarrow }^\dag a_{R \downarrow }^\dag {a_{R \downarrow }}{a_{R \downarrow }} \\
 \end{array} \right)\\
 {H_{ \uparrow  \downarrow }} &=& {U_2}\left( {a_{L \uparrow }^\dag a_{L \downarrow }^\dag {a_{L \uparrow }}{a_{L \downarrow }} + a_{R \uparrow }^\dag a_{R \downarrow }^\dag {a_{R \uparrow }}{a_{R \downarrow }}} \right)
\end{eqnarray}
$a^\dag$ and $a$ are respectively the bosonic creation and annihilation operators for the polariton or exciton field, the subscripts $L,R$ and $\uparrow,\downarrow$ refer respectively to condensed particles in the left or right trap with $\sigma_+$ or $\sigma_-$ polarization. $H_0$ is the free particles Hamiltonian, $H_J$ stands for the spin conservative Josephson tunneling, $H_\Omega$ models the spin flip process induced by the structural anisotropy and $H_{pol - pol}$ is the condensed particles interaction term which contains respectively parallel ($H_{\uparrow\uparrow }$) and antiparallel ($H_{\uparrow\downarrow }$) spin scattering processes.

And the decoherent part:
\begin{equation}\label{HpHm}
{H_{deco}} = {H_ + } + {H_ - }
\end{equation}
with:
\begin{eqnarray}
{H_ + } &=& D\sum\limits_{i = 1}^N {\left\{ \begin{array}{l}
a_{L \uparrow }^\dag {a_{i \uparrow }}b_i^\dag  + a_{L \downarrow }^\dag {a_{i \downarrow }}b_i^\dag  \\
+ a_{R \uparrow }^\dag {a_{i \uparrow }}b_i^\dag  + a_{R \downarrow }^\dag {a_{i \downarrow }}b_i^\dag  \\
\end{array} \right\}} \label{Hp} \\
{H_ - } &=& D\sum\limits_{i = 1}^N {\left\{ \begin{array}{l}
{a_{L \uparrow }}a_{i \uparrow }^\dag {b_i} + {a_{L \downarrow }}a_{i \downarrow }^\dag {b_i} \\
+{a_{R \uparrow }}a_{i \uparrow }^\dag {b_i} + {a_{R \downarrow }}a_{i \downarrow }^\dag {b_i}
\end{array} \right\}}\label{Hm}
\end{eqnarray}
which describes the spin conserving interactions between condensed particles and the acoustic phonons reservoir described by $b_i^\dag$ and $b_i$ operators. The decoherent Hamiltonian is split the following way: $H_-$ models the excitation of a L or R condensed particle toward the $i^{th}$ of the $N$ trapped state ($a_i^\dag$ operator) via the absorbtion of a phonon with energy $\hbar\omega_i$ and $H_+$ represents the opposite relaxation scheme. We describe the dynamic of the system by means of density matrix formalism.

The time evolution under the coherent part of the Hamiltonian is treated by the usual Liouville-von Neumann equation:
\begin{equation}\label{Liouville}
{\left( {{\partial _t}\rho } \right)_{co}} = \frac{i}{\hbar }\left[ {\rho;{H_{co}}} \right]
\end{equation}
which gives the following equation of motion for mean value of an arbitrary operator $\langle \widehat{A}\rangle=Tr(\rho\widehat{A})$:
\begin{equation}\label{dtco}
    {\partial _t}{\left\langle {\widehat A} \right\rangle _{co}} = \frac{i}{\hbar }Tr\left( {\left[ {\rho ,{H_{co}}} \right]\widehat A} \right)=\frac{i}{\hbar }\left\langle {\left[ {{H_{co}},\widehat A} \right]} \right\rangle
\end{equation}

Applying this formula to the operators of the number of spin- up and spin down polaritons in right and left wells ${n_{\left\{ {L,R} \right\}\left\{ { \uparrow , \downarrow } \right\}}}=a_{\left\{ {L,R} \right\}\left\{ { \uparrow , \downarrow } \right\}}^\dag {a_{\left\{ {L,R} \right\}\left\{ { \uparrow , \downarrow } \right\}}}$ and to the correlators giving the orientation of the linear polarization  ${a_{\left\{ {R,L} \right\} \uparrow }^\dag a_{\left\{ {R,L} \right\} \downarrow } }$, ${a_{\left\{ {R,L} \right\} \uparrow }^\dag a_{\left\{ {L,R} \right\} \downarrow } }$ and ${a_{\left\{ {R,L} \right\} \uparrow }^\dag a_{\left\{ {L,R} \right\} \uparrow } }$ and using mean field approximation for truncation of the fourth- order correlators one can get a closed set of ten evolution equations. In a matter of saving space we will only write one sample of each type of equation, the remaining complementary ones are straightforwardly obtained by permutations of $L,R$ and $\uparrow,\downarrow$ indexes. We also compact the notation for mean values of correlators ie for example $\left\langle {a_{R \downarrow }^\dag {a_{L \uparrow }}} \right\rangle$ becomes $\alpha _{RL}^{ \downarrow  \uparrow }$:
\begin{widetext}
\begin{eqnarray}
{\hbar\left( {{\partial _t}{n_{L \uparrow }}} \right)_{co}} =  - 2J{\mathop{\rm Im}\nolimits} \left( {\alpha _{RL}^{ \uparrow  \uparrow }} \right) - 2\Omega {\mathop{\rm Im}\nolimits} \left( {\alpha _{LL}^{ \downarrow  \uparrow }} \right)
\end{eqnarray}
\begin{eqnarray}
{\hbar\left( {{\partial _t}\alpha _{LL}^{ \downarrow  \uparrow }} \right)_{co}} = iJ\left[ {\alpha _{RL}^{ \downarrow  \uparrow } - \alpha {{_{RL}^{ \uparrow  \downarrow *}}}} \right] + i\Omega \left[ {{n_{L \uparrow }} - {n_{L \downarrow }}} \right] + i\left[ {{U_1}\left( {{n_{L \downarrow }} - {n_{L \uparrow }} } \right) + {U_2}\left( {{n_{L \uparrow }} - {n_{L \downarrow }} } \right)} \right]\alpha _{LL}^{ \downarrow  \uparrow }
\end{eqnarray}
\begin{eqnarray}
{\hbar\left( {{\partial _t}\alpha _{RL}^{ \uparrow  \uparrow }} \right)_{co}} = iJ\left[ {{n_{L \uparrow }} - {n_{R \uparrow }}} \right] + i\Omega \left[ {\alpha _{RL}^{ \downarrow  \uparrow } - \alpha _{RL}^{ \uparrow  \downarrow }} \right] + i\left[ {{U_1}\left( {{n_{R \uparrow }} - {n_{L \uparrow }} } \right) + {U_2}\left( {{n_{R \downarrow }} - {n_{L \downarrow }}} \right)} \right]\alpha _{RL}^{ \uparrow  \uparrow }
\end{eqnarray}
\begin{eqnarray}
{\hbar\left( {{\partial _t}\alpha _{RL}^{ \downarrow  \uparrow }} \right)_{co}} = iJ\left[ {\alpha _{LL}^{ \downarrow  \uparrow } - \alpha _{RR}^{ \downarrow  \uparrow }} \right] + i\Omega \left[ {\alpha _{RL}^{ \uparrow  \uparrow } - \alpha _{RL}^{ \downarrow  \downarrow }} \right] + i\left[ {{U_1}\left( {{n_{R \downarrow }} - {n_{L \uparrow }} } \right) + {U_2}\left( {{n_{R \uparrow }} - {n_{L \downarrow }}} \right)} \right]\alpha _{RL}^{ \downarrow  \uparrow }
\end{eqnarray}
\end{widetext}

The dynamics of the decoherent part involving phonons is dissipative, and should be treated in the following way. Liouville- von Neumann equation can be rewritten in a following integro-differential form:
\begin{equation}\label{LVNint}
{\left( {{\partial _t}\rho } \right)_{deco}} =  - \frac{1}{\hbar^2}\int\limits_{ - \infty }^t {\left[ {{H_{deco}}\left( t \right);\left[ {{H_{deco}}\left( {t'} \right);\rho \left( {t'} \right)} \right]} \right]dt'}
\end{equation}
where $H_{deco}(t)$ is the time- dependent Hamiltonian of the polariton- phonon interaction \ref{HpHm} written in a Dirac picture. To account for the decoherent nature of the evolution with phonons, Born-Markov approximation should be applied while treating Eq.\ref{LVNint}. It consists in replacing $t'$ by $t$ which retains only energy- conserving terms after integration \cite{Carmichael}. The time evolution of the density matrix considering Eq.\ref{HpHm} is thus given by the following master equation:

\begin{eqnarray} \delta^{-1}(\Delta E)\hbar\partial_t\rho &=& 2\left(H_+\rho H_-+H_-\rho H_+\right)\\
\nonumber &-&\left(H_+H_-+H_-H_+\right)\rho\\
\nonumber &-&\rho\left(H_+H_-+H_-H_+\right)
\label{Linblad}
\end{eqnarray}
where the factor $\delta^{-1}(\Delta E)$ denotes the conservation of
energy. For time evolution of the mean value of any arbitrary operator $\widehat{A}$ one has:
\begin{eqnarray}\label{eqM}
\delta^{-1}(\Delta E)\hbar\left(\partial_t\langle \widehat{A}\rangle\right)_{deco}&=&Tr\left(\rho[H_-;[\widehat{A};H_+]]\right)\\
\nonumber &+&Tr\left(\rho[H_+;[\widehat{A};H_-]]\right)
\end{eqnarray}
We now apply Eq.\ref{eqM} to the previous densities and correlators with Eqs.(\ref{Hp}-\ref{Hm}) to obtain the set of equations for the decoherent part, and once again we only write the six foretype equations:
\begin{widetext}
\begin{eqnarray}
{\left( {{\partial _t}{n_{L \uparrow }}} \right)_{deco}} = 2{W}\sum\limits_{i = 1}^N {{\mathop{\rm Re}\nolimits} \left\{ \begin{array}{l}
 \left( {n_i^{ph} + 1} \right)\left[ {{n_{i \uparrow }}\left( {{n_{L \uparrow }} + 1 + \alpha _{RL}^{ \uparrow  \uparrow *}} \right) + \alpha _{ii}^{ \downarrow  \uparrow }\left( {\alpha _{LL}^{ \downarrow  \uparrow *} + \alpha _{RL}^{ \downarrow  \uparrow *}} \right)} \right] \\
  - n_i^{ph}\left[ {\left( {{n_{i \uparrow }} + 1} \right)\left( {{n_{L \uparrow }} + \alpha _{RL}^{ \uparrow  \uparrow *}} \right) + \alpha _{ii}^{ \downarrow  \uparrow }\left( {\alpha _{LL}^{ \downarrow  \uparrow *} + \alpha _{RL}^{ \downarrow  \uparrow *}} \right)} \right] \\
 \end{array} \right\}}
\end{eqnarray}
\begin{eqnarray}
{\left( {{\partial _t}\alpha _{LL}^{ \downarrow  \uparrow }} \right)_{deco}} = {W}\sum\limits_{i = 1}^N {\left\{ \begin{array}{l}
 \left( {n_i^{ph} + 1} \right)\left[ \begin{array}{l}
 \left( {{n_{i \downarrow }} + {n_{i \uparrow }}} \right)\alpha _{LL}^{ \downarrow  \uparrow } + {n_{i \downarrow }}\alpha _{RL}^{ \downarrow  \uparrow } + {n_{i \uparrow }}\alpha _{RL}^{ \uparrow  \downarrow *} \\
  + \alpha _{ii}^{ \downarrow  \uparrow }\left( {{n_{L \downarrow }} + {n_{L \uparrow }} + 2 + \alpha _{RL}^{ \uparrow  \uparrow } + \alpha _{RL}^{ \downarrow  \downarrow *}} \right) \\
 \end{array} \right] \\
  - n_i^{ph}\left[ \begin{array}{l}
 \left( {{n_{i \downarrow }} + {n_{i \uparrow }} + 2} \right)\alpha _{LL}^{ \downarrow  \uparrow } + \left( {{n_{i \downarrow }} + 1} \right)\alpha _{RL}^{ \downarrow  \uparrow } \\
  + \left( {{n_{i \uparrow }} + 1} \right)\alpha _{RL}^{ \uparrow  \downarrow *} + \alpha _{ii}^{ \downarrow  \uparrow }\left( {{n_{L \downarrow }} + {n_{L \uparrow }} + \alpha _{RL}^{ \uparrow  \uparrow } + \alpha _{RL}^{ \downarrow  \downarrow *}} \right) \\
 \end{array} \right] \\
 \end{array} \right\}}
\end{eqnarray}
\begin{eqnarray}
{\left( {{\partial _t}\alpha _{RL}^{ \uparrow  \uparrow }} \right)_{deco}} = W\sum\limits_{i = 1}^N {\left\{ \begin{array}{l}
 \left( {n_i^{ph} + 1} \right)\left[ \begin{array}{l}
 {n_{i \uparrow }}\left( {{n_{L \uparrow }} + {n_{R \uparrow }} + 2 + 2\alpha _{RL}^{ \uparrow  \uparrow }} \right) \\
  + \alpha _{ii}^{ \downarrow  \uparrow }\left( {\alpha _{RL}^{ \uparrow  \downarrow } + \alpha _{RR}^{ \downarrow  \uparrow *}} \right) + \alpha _{ii}^{ \downarrow  \uparrow *}\left( {\alpha _{RL}^{ \downarrow  \uparrow } + \alpha _{LL}^{ \downarrow  \uparrow }} \right) \\
 \end{array} \right] \\
  - n_i^{ph}\left[ \begin{array}{l}
 \left( {{n_{i \uparrow }} + 1} \right)\left( {{n_{L \uparrow }} + {n_{R \uparrow }} + 2\alpha _{RL}^{ \uparrow  \uparrow }} \right) \\
  + \alpha _{ii}^{ \downarrow  \uparrow }\left( {\alpha _{RL}^{ \uparrow  \downarrow } + \alpha _{RR}^{ \downarrow  \uparrow *}} \right) + \alpha _{ii}^{ \downarrow  \uparrow *}\left( {\alpha _{RL}^{ \downarrow  \uparrow } + \alpha _{LL}^{ \downarrow  \uparrow }} \right) \\
 \end{array} \right] \\
 \end{array} \right\}}
\end{eqnarray}
\begin{eqnarray}
{\left( {{\partial _t}\alpha _{RL}^{ \downarrow  \uparrow }} \right)_{deco}} = {W}\sum\limits_{i = 1}^N {\left\{ \begin{array}{l}
 \left( {n_i^{ph} + 1} \right)\left[ \begin{array}{l}
 {n_{i \downarrow }}\left( {\alpha _{LL}^{ \downarrow  \uparrow } + \alpha _{RL}^{ \downarrow  \uparrow }} \right) + {n_{i \uparrow }}\left( {\alpha _{RR}^{ \downarrow  \uparrow } + \alpha _{RL}^{ \downarrow  \uparrow }} \right) \\
  + \alpha _{ii}^{ \downarrow  \uparrow }\left( {{n_{R \downarrow }} + {n_{L \uparrow }} + 2 + \alpha _{RL}^{ \uparrow  \uparrow } + \alpha _{RL}^{ \downarrow  \downarrow }} \right) \\
 \end{array} \right] \\
  - n_i^{ph}\left[ \begin{array}{l}
 \left( {{n_{i \downarrow }} + 1} \right)\left( {\alpha _{LL}^{ \downarrow  \uparrow } + \alpha _{RL}^{ \downarrow  \uparrow }} \right) + \left( {{n_{i \uparrow }} + 1} \right)\left( {\alpha _{RR}^{ \downarrow  \uparrow } + \alpha _{RL}^{ \downarrow  \uparrow }} \right) \\
  + \alpha _{ii}^{ \downarrow  \uparrow }\left( {{n_{R \downarrow }} + {n_{L \uparrow }} + \alpha _{RL}^{ \uparrow  \uparrow } + \alpha _{RL}^{ \downarrow  \downarrow }} \right) \\
 \end{array} \right] \\
 \end{array} \right\}}
\end{eqnarray}

\begin{eqnarray}
{\left( {{\partial _t}{n_{i \uparrow }}} \right)_{deco}} = 2{W}\sum\limits_i {{\mathop{\rm Re}\nolimits} \left\{ \begin{array}{l}
 n_i^{ph}\left[ \begin{array}{l}
 \left( {{n_{i \uparrow }} + 1} \right)\left( {{n_{L \uparrow }} + {n_{R \uparrow }} + 2{\mathop{\rm Re}\nolimits} \left( {\alpha _{RL}^{ \uparrow  \uparrow }} \right)} \right) \\
  + \alpha _{ii}^{ \downarrow  \uparrow }\left( {\alpha _{LL}^{ \downarrow  \uparrow *} + \alpha _{RR}^{ \downarrow  \uparrow *} + \alpha _{RL}^{ \downarrow  \uparrow *} + \alpha _{RL}^{ \uparrow  \downarrow }} \right) \\
 \end{array} \right] \\
  - \left( {n_i^{ph} + 1} \right)\left[ \begin{array}{l}
 {n_{i \uparrow }}\left( {{n_{L \uparrow }} + {n_{R \uparrow }} + 2 + 2{\mathop{\rm Re}\nolimits} \left( {\alpha _{RL}^{ \uparrow  \uparrow }} \right)} \right) \\
  + \alpha _{ii}^{ \downarrow  \uparrow }\left( {\alpha _{LL}^{ \downarrow  \uparrow *} + \alpha _{RR}^{ \downarrow  \uparrow *} + \alpha _{RL}^{ \downarrow  \uparrow *} + \alpha _{RL}^{ \uparrow  \downarrow }} \right) \\
 \end{array} \right] \\
 \end{array} \right\}}
\end{eqnarray}
\begin{eqnarray}
{\left( {{\partial _t}\alpha _{ii}^{ \downarrow  \uparrow }} \right)_{deco}} = {W}\sum\limits_i {\left\{ \begin{array}{l}
 n_i^{ph}\left[ \begin{array}{l}
 \left( {{n_{i \downarrow }} + {n_{i \uparrow }} + 2} \right)\left( {\alpha _{LL}^{ \downarrow  \uparrow } + \alpha _{RR}^{ \downarrow  \uparrow } + \alpha _{RL}^{ \uparrow  \downarrow *} + \alpha _{RL}^{ \downarrow  \uparrow }} \right) \\
  + \alpha _{ii}^{ \downarrow  \uparrow }\left( {{n_{L \uparrow }} + {n_{L \downarrow }} + {n_{R \uparrow }} + {n_{R \downarrow }} + 2{\mathop{\rm Re}\nolimits} \left( {\alpha _{RL}^{ \uparrow  \uparrow } + \alpha _{RL}^{ \downarrow  \downarrow }} \right)} \right) \\
 \end{array} \right] \\
  - \left( {n_i^{ph} + 1} \right)\left[ \begin{array}{l}
 \left( {{n_{i \downarrow }} + {n_{i \uparrow }}} \right)\left( {\alpha _{LL}^{ \downarrow  \uparrow } + \alpha _{RR}^{ \downarrow  \uparrow } + \alpha _{RL}^{ \uparrow  \downarrow *} + \alpha _{RL}^{ \downarrow  \uparrow }} \right) \\
  + \alpha _{ii}^{ \downarrow  \uparrow }\left( {{n_{L \uparrow }} + {n_{L \downarrow }} + {n_{R \uparrow }} + {n_{R \downarrow }} + 4 + 2{\mathop{\rm Re}\nolimits} \left( {\alpha _{RL}^{ \uparrow  \uparrow } + \alpha _{RL}^{ \downarrow  \downarrow }} \right)} \right) \\
 \end{array} \right] \\
 \end{array} \right\}}
\end{eqnarray}
\end{widetext}
We get $2N$ extra equations for the excited levels populations ${{n_{i \uparrow }}}$ and ${{n_{i \downarrow }}}$ plus N more for the new associated correlators ${\alpha _{ii}^{ \downarrow  \uparrow }}$ so a total of (10+3N) equations which only allow numerical considerations. The scattering rates can be estimated as $W=D^2/\gamma\hbar$, with $\gamma$ being a characteristic broadening of the level, $n_i^{ph}=b_i^\dag {b_i}$ is the distribution of phonons with energy $E_i=\hbar\omega_i$ at the temperature $T$ given by Bose-Einstein distribution.

In order to get closer to a realistic situation, one can include in the model the phonon- assisted inter-excited levels transitions processes by adding to the decoherent Hamiltonian (Eq.\ref{HpHm}) the extra term:
\begin{equation}
{H_{int-ex}} = D\left\{ \begin{array}{l}
 \sum\limits_{i < j}^N {\left( {{a_{i \uparrow }}a_{j \uparrow }^\dag {b_{j - i}} + {a_{i \downarrow }}a_{j \downarrow }^\dag {b_{j - i}}} \right)}  \\
  + \sum\limits_{i > j}^N {\left( {a_{j \uparrow }^\dag {a_{i \uparrow }}b_{i - j}^\dag  + a_{j \downarrow }^\dag {a_{i \downarrow }}b_{i - j}^\dag } \right)}  \\
 \end{array} \right\}\label{Hintex}
\end{equation}
which participates to time dependent equations the following way (the 2 spin down equations are obviously the same):
\begin{widetext}
\begin{eqnarray}
 \left( {{\partial _t}{n_{i > j \uparrow }}} \right)_{deco}^{int - ex} &=& {W}\sum\limits_{i > j}^N {\left\{ { + n_{i - j}^{ph}\left( {{E_{i - j}},T} \right){n_{j \uparrow }} - \left( {1 + n_{i - j}^{ph}\left( {{E_{i - j}},T} \right) + {n_{j \uparrow }}} \right){n_{i \uparrow }}} \right\}}  \\
 \left( {{\partial _t}{n_{i < j \uparrow }}} \right)_{deco}^{int - ex} &=& {W}\sum\limits_{i < j}^N {\left\{ { - n_{j - i}^{ph}\left( {{E_{j - i}},T} \right){n_{i \uparrow }} + \left( {1 + n_{j - i}^{ph}\left( {{E_{j - i}},T} \right) + {n_{i \uparrow }}} \right){n_{j \uparrow }}} \right\}}
\end{eqnarray}
\end{widetext}

The full set of equations for the dynamics of the system is finally obtained by:
\begin{equation}\label{Final}
{\partial _t}\left\langle {\widehat{A}} \right\rangle  = {\partial _t}{\left\langle {\widehat{A}} \right\rangle _{co}} + {\partial _t}{\left\langle {\widehat{A}} \right\rangle _{deco}} + {\partial _t}\left\langle {\widehat{A}} \right\rangle _{deco}^{int - ex}
\end{equation}
We are then able to describe the long living indirect excitonic system. Now, to study properly the behavior of the decaying polaritonic condensate, one also has to include pumping and lifetime to the Hamiltonian. The last results of the numerical
experiment described in the next section involve non-resonant pumping of the excited levels, which can be easily introduced in the master equation \cite{Carmichael} and give standard pump and decay terms in dynamic equations for the occupancies. The addition to all equations due to finite lifetime is
\begin{equation}\label{lifetime}
{\partial _t}\left\langle {\widehat{A}} \right\rangle _{lt} = -\left\langle {\widehat{A}} \right\rangle/\tau
\end{equation}
where $\tau$ is the particle lifetime. For the pumping, a constant rate term is added only to the equations for the populations
\begin{equation}\label{pump}
{\partial _t}\left(n_\nu \right) _{pump} =  P_\nu
\end{equation}

\section{Results and discussion}

In this section we present the data of the numerical simulations of the system of the kinetic equations derived in the previous section.
In all calculations, the values chosen for the parameters when they are not taken to be zero are $J=50$ $\mu$eV, $\Omega=60$ $\mu$eV, $U=3$ $\mu$eV, $W=2.23\cdot10^{9}s^{-1}$ and $T=20$ $K$.

\subsection{Josephson oscillations and self-trapping}

We start by considering the case of particles of infinite lifetime, corresponding to indirect excitons in coupled quantum wells. There are therefore no decay or pumping terms included and we start from some hypothetical initial population of the ground states, the excited states being initially unpopulated. In the third subsection of the results, finite particle lifetime is introduced along with pumping, corresponding to the polariton system.

\begin{figure}[tbp]
\includegraphics[width=1.0\linewidth]{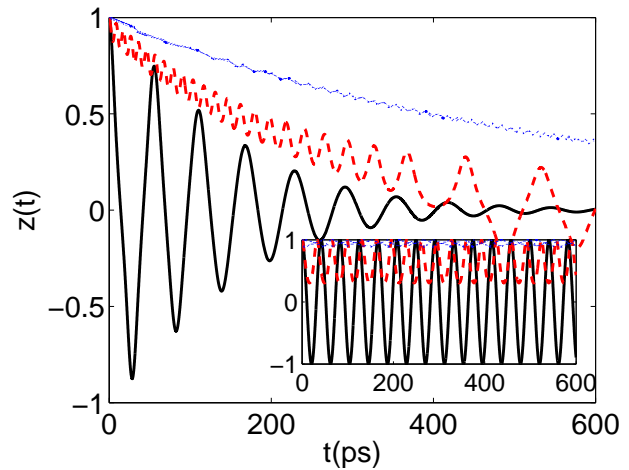}
\caption{Josephson oscillations for the case $\tau=0$. Spin degree of freedom is neglected to emphasize effects of polariton-polariton and polariton-phonon interactions. Curves show population imbalance at different occupation numbers.(Solid/black: $N=100$, dashed/red: $N=200$, dotted/blue: $N=500$) Inset: the oscillations for the case when interaction with phonons are absent.(Solid/black: $N=10$, dashed/red: $N=70$, dotted/blue: $N=150$) One sees that polariton- polariton interactions lead to self- trapping effect, shortened period and anharmonicity, while interaction with phonons lead to the damping of the Josephson oscillations and increased population needed to reach self-trapping.}
\label{selftrapping}
\end{figure}

 Fig.\ref{selftrapping} shows the Josephson oscillations, when all particles are initially in the left well. We neglected the spin degree of freedom for simplicity and to emphasize the effect of exciton-exciton and exciton-phonon interactions. The figure shows the time evolution of the population imbalance, defined as $$z=\frac{n_L-n_R}{n_L+n_R}$$
The inset shows the oscillations in the absence of phonons. The main plot shows that the phonon interactions causes a damping of the oscillations compared to the inset plot. The effect of increasing the population and therefore exciton-exciton scattering, is the self-trapping of the particles, shortened period of oscillations and anharmonicity which is very obvious in the time interval $400-600$ ps for the dashed/red curve. Another observation is that the phonon damping decreases the effect of exciton-exciton scattering in the sense that much higher population is needed to get the self-trapping.

\subsection{Spatial separation of polarization}
\begin{figure}[tbp]
\includegraphics[width=1.0\linewidth]{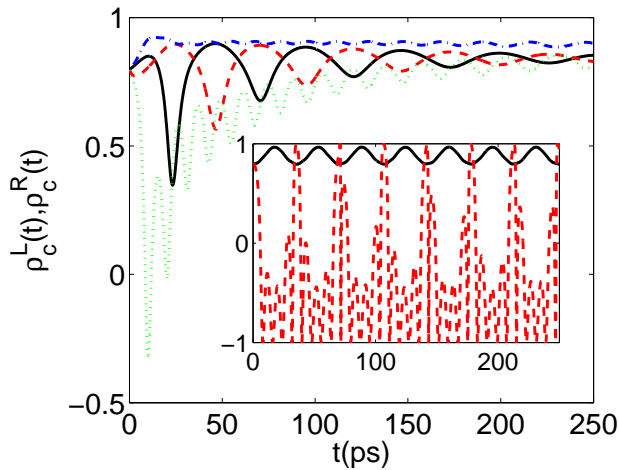}
\caption{Temporal dependence of circular polarization degree in both traps. We let $\Omega=0$ for simplicity. Initially there are no polaritons in the right trap, N in the left with initial circular polarization degree $\rho_c=0.8$. ($N=100$: Solid/black: L, dashed/red: R; $N=200$: dash-dot/blue: L, dotted/green: R) For $N=100$, no self-trapping occurs and the L/R polarizations oscillate in a similar way. For $N=200$, the spin-up particles are self-trapped and the system shows a spatial separation of polarizations. Inset: $N=200$, no phonons. One sees a similar separation of polarizations, but the polarization of the right trap goes to 1 periodically as the spin-down component is emptied.}
\label{polarizations}
\end{figure}

Accounting for the spin degree of freedom, one can observe intriguing phenomena in the polarization domain.  We let $\Omega=0$ (no spin flips) for the moment to simplify the interpreting of the results. In this case, the spin up component behaves quite independently of the spin down component. They do affect each other slightly through the exciton-exciton scattering, but as already stated, antiparallel spins usually interact much weaker than parallel ones. We can thus have three situations, where both spin components are self-trapped, only one of them, or neither. Fig.\ref{polarizations} shows the behavior of the polarization degree in both traps for two different population numbers but same initial polarization degree. Initially all particles are in the left trap. The solid/black and the dashed/red curves are the circular polarization degrees in the left and right traps, respectively, for $N=100$. Neither spin component is self-trapped, so the oscillations of the polarizations are quite similar. For $N=200$ (dash-dot/blue: L, dotted/green: R) the spin-up component is self-trapped, and we get a spatial separation of polarization. The oscillations are damped by the phonon interaction, and for large times the polarizations equalize. The inset shows the case of $N=200$ without phonon interaction. There is a visible separation of the polarizations, but without damping, the spin-down component makes full oscillations between the two traps and thus the polarization degree periodically reaches $1$. The short oscillations in the right trap polarization (dashed/red) are due to the oscillations of the spin-up component, which are not visible in the left trap polarization.

\subsection{Bistability and sustained oscillations}
For a consistent description of Josephson tunneling in polariton systems one needs to introduce pumping and decay terms into kinetic equations. In the case where decoherence in the system due to the interaction with acoustic phonons is neglected, the dynamics can be described by a driven dissipative Gross-Pitaevskii equation \cite{Gippius2007}, which in some range of parameters can have several different stationary solutions. The system thus demonstrates multistability as recently observed experimentally \cite{Naturematerials2010}. In the case of the quasi-resonant pumping, the bi(multi)-stability occurs when the pumping laser lies above the energy of the interacting polariton state. An increase of the pumping results in a larger population of the state which becomes closer in energy to the one of the laser, increasing the light absorption. Above some threshold, an avalanche effect takes place and the system jumps to a new stability point with a much larger population and an energy of the state lying above the one of the laser. Decreasing the pumping intensity from this higher stability branch the system will jump back to the branch corresponding to a smaller population only for a pumping much below the one allowing to jump up, leading to the formation an hysteresis cycle. Although in a present paper we use a different approach than in \cite{Gippius2007} for the description of the polariton system, one can expect that a similar type of effect will appear in our results. In our model the low population stable branch corresponds to the case a balanced population in the two wells. The high population stable branch corresponds to the self trapping case.

\begin{figure}[tbp]
\includegraphics[width=1.0\linewidth]{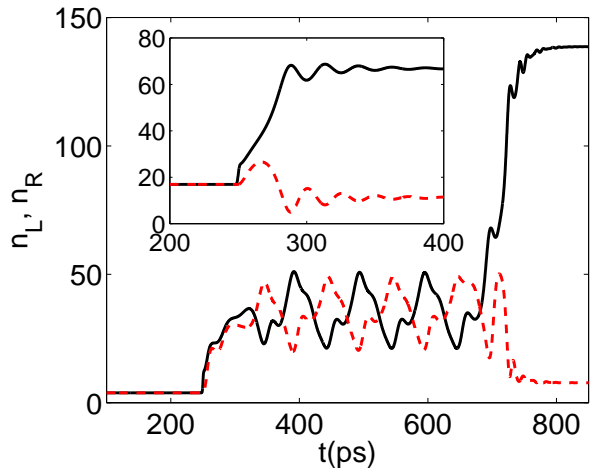}
\caption{Bistability in polariton Josephson junction with pump and decay. Main figure: $\Omega=60$ $\mu$eV. At $t=250$ ps and $t=650$ ps the amplitude of the constant pump increases in a step-like manner. In the intermediate regime of moderate pumping, the system reveals self- sustained oscillations. (A short probe was sent to the left trap at $t=250$ ps to break the symmetry.)
Inset: $\Omega=0$, spin degree of freedom neglected. In this case, no self-sustained oscillations can be observed, but the bistability jump remains. A short probe is sent to the left trap at the moment of the pump jump. }
\label{bistability}
\end{figure}

Fig.\ref{bistability} shows the behavior of polaritons with a lifetime of 16 ps. In both plots, the spin-down component of the first excited level is being pumped continuously at varying pumping strengths. As the excited levels are delocalized, the non-resonant pumping is spatially homogeneous.
The inset shows the case where only one spin component is considered. Up until $t=250$ ps, the excited level is being pumped with a low strength, and the equilibrium state is stable. At $t=250$ ps the pumping strength is increased beyond some threshold which causes the populations to split and stabilize in a state where the occupancy of one of the traps is much higher than the other. In numerical calculations, this requires sending a very small asymmetric probe to get out of the unstable equilibrium.

\begin{figure}[tbp]
\begin{tabular}{c}
\includegraphics[width=1.0\linewidth]{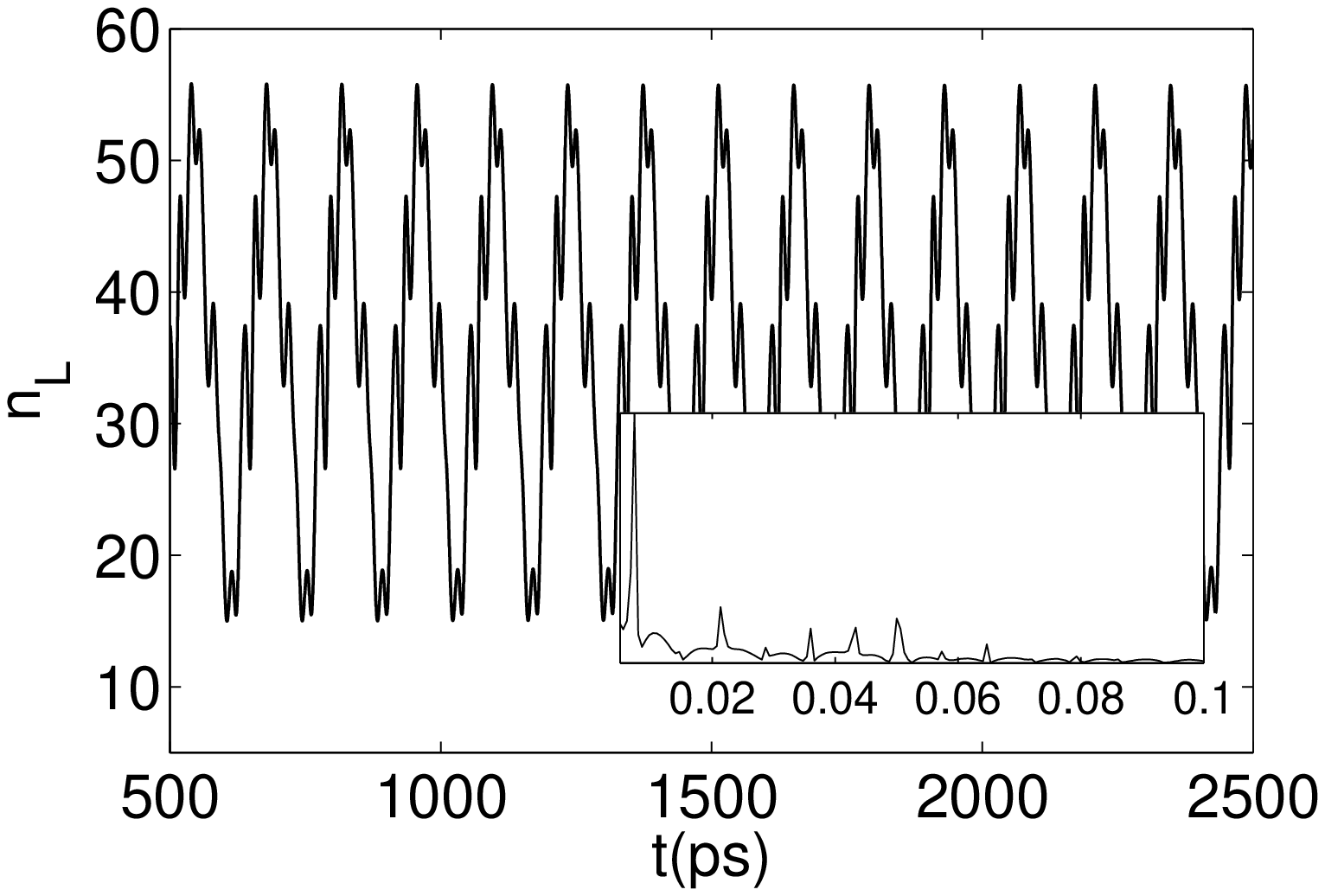}\\
\includegraphics[width=1.0\linewidth]{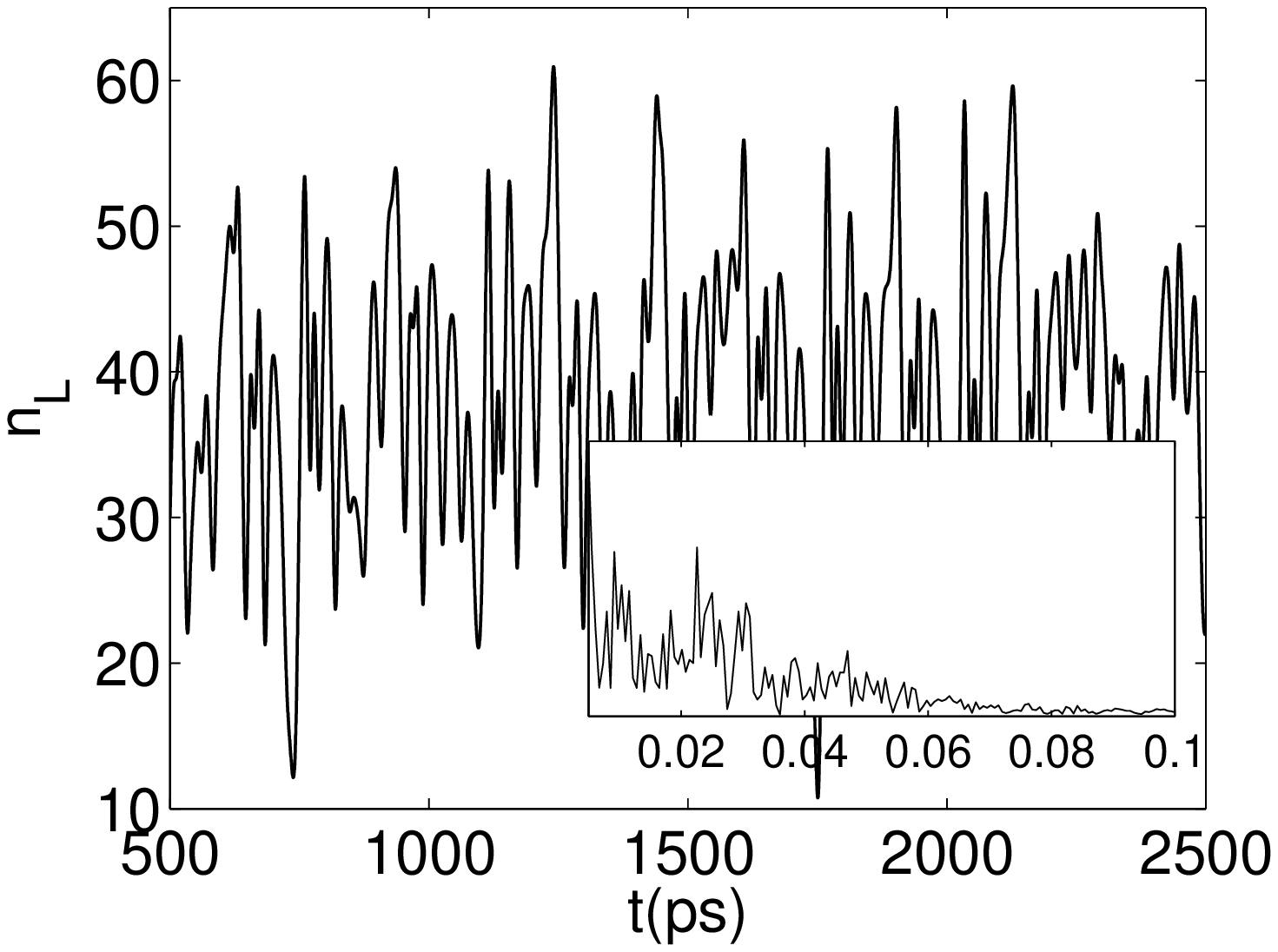}
\end{tabular}
\caption{Two plots showing the difference between chaotic and non-chaotic oscillations in the system. In the upper plot, $\Omega=55$ eV and $P=10$ ps$^{-1}$, and the oscillations are non-chaotic, characterized by a a Fourier spectrum with sharp peaks. The inset shows the Fourier spectrum in arbitrary amplitude units as a function of frequency [ps$^{-1}$]. In the lower plot, $\Omega=90$ eV and $P=13$ ps$^{-1}$. Here, the oscillations are chaotic.}
\label{chaotic}
\end{figure}

The situation gets even more interesting when one considers both spins and introduces the coherent coupling  between the two. In this case, there is also a stable state of split populations for high pumping strength and equal populations for low pumping strength, but in between there is a regime where the populations do not reach stationary values, but reveals self- sustained oscillations resulting from the interplay between intrinsic and extrinsic Josephson effects as is shown in the main plot of Fig.\ref{bistability}. (Only the spin-down particles are shown for readability) This can be explained in terms of a Hopf bifurcation appearing in our system, where for a certain range of parameters the equilibrium point becomes unstable and instead a stable limit cycle is created. Another interesting effect is that the oscillations are not necessarily periodic and become chaotic for a certain range of parameters, which means that strange attractor instead of a limit cycle is formed in the system. The oscillations and their Fourier spectra are shown for two cases in Fig.\ref{chaotic}. They are for $\Omega=55,90$ eV and pumping strengths $P=10, 13$ ps$^{-1}$, respectively. In the first case, the oscillations are periodic, as characterized by their Fourier spectrum with sharp peaks. The second case shows chaotic oscillations with a broadband Fourier spectrum. These periodic and chaotic oscillations have previously been predicted for a polaritonic system without dissipation in \cite{Solnyshkov2009}.

\begin{figure}[tbp]
\includegraphics[width=1.0\linewidth]{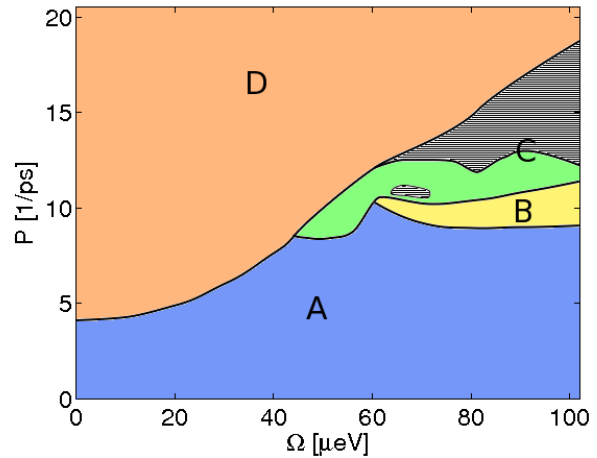}
\caption{Phase diagram for polariton system with constant non- resonant circularly polarized pumping of one of the excited levels. A: Stable state of equal left and right populations. Slight spin splitting because of the circularly polarized pumping. B: Spin-up and spin-down particles enter bistable states separately. C: Self-sustained oscillations. Shaded parts are where chaotic oscillations can be observed. D: Massive splitting where one of the populations with same polarization as the pump reaches a high value while the others have a low value. }
\label{phases}
\end{figure}

Fig.\ref{phases} shows a phase diagram of the system in axes of pumping strength $P$ vs. the spin coupling $\Omega$. The pump is still applied to the spin-down component of the first delocalized excited level. The area A corresponds to low pumping, where the only population splitting is the one between the spin up and spin down. In B, both spin components get split equally. C is the range of parameters which give rise to sustained oscillations, and in D there is a massive split-off of one of the spin-down populations while spin-up populations remain low. The shaded part of C is where chaotic oscillations can be observed.

\section{Conclusions}

In conclusion, we analyzed the Josephson- related phenomena in coupled condensates of indirect excitons and cavity polaritons taking into account their peculiar spin structure, particle- particle interactions, interactions with phonons and pump and decay terms. For indirect excitons having extremely long lifetimes we have shown that exciton- exciton interactions lead to anharmonicity of Josephson oscillations, self trapping effect and spontaneous separation of the fractions with opposite circular polarizations in the real space. The main effect of the particle-phonon interaction is to dampen the oscillations and raise the population threshold for the self-trapping to appear. For cavity polaritons having short lifetimes we demonstrated the bistable behavior of the Josephson junction in the regime of the incoherent constant pump. We have shown that the account of the coupling between the polaritons with opposite circular polarizations can qualitatively change the bistability pattern and in some range of the parameters lead to self- sustained oscillations. These oscillations can then have periodic or chaotic behaviour.

The authors thank D.D. Solnyshkov and G. Pavlovic for useful discussions. EBM and IAS acknowledge the support from FP7 IRSES project "SPINMET" and from Rannis "Center of excellence in polaritonics". The authors aknowlegde the support from the FP7 project Spin-Optronics, Grant Agreement 237252 and from joint French- Icelandic Project Jules Verne.


\begin{thebibliography}{99}
\bibitem{DarkSpinRemark} The same is valid for indirect excitons.
\end{thebibliography}
\end{document}